\documentclass[twocolumn,showpacs,preprintnumbers,amsmath,amssymb]{revtex4}

\usepackage{graphicx}
\usepackage{dcolumn}
\usepackage{bm}

\begin{document}


\title{{\it Ab initio} study of the influence of adsorbed atoms on vacancy-induced 
magnetic moments in graphene sheets}

\author{Rodrigo Yoshikawa Oeiras}
\author{Fernando M. Ara\'ujo-Moreira}
\affiliation{Departamento  de  F\'{\i}sica,  Universidade  Federal  de
S\~ao Carlos, CP 676, S\~ao Carlos, SP, Brazil, CEP 13565-905}

\author{Marcos  Ver\'{\i}ssimo-Alves}
\email{mverissi@ictp.it}
\affiliation{The Abdus Salam International Centre for Theoretical Physics, Strada
Costiera 11, Main Building, Trieste I-34014, Italy}

\author{Ricardo Faccio, Helena Pardo and Alvaro W. Mombr\'u}

\affiliation{Crystallography, Solid State and Materials Laboratory (Cryssmat-Lab),
Universidad de la Rep\'ublica, P.O. Box 1157, CP 11800, Montevideo, Uruguay}

\date\today

\begin{abstract}
We present {\it ab initio} calculations for single-atom vacancies in
graphene. In agreement with earlier work, we find that vacancies are
responsible for the magnetism recently observed experimentally,
creating a center with net spin. For small supercells,
there is a strong symmetry breaking in the supercell with respect to
the perfect graphene structure, and this symmetry breaking is much
smaller for larger supercells. The influence of 
adsorption of H, O and N atoms on the spin center of the bare 
single-atom vacancy and its influence
on the magnetism induced in these samples is investigated. A rich variety of mechanisms
is found for the generation of magnetic moments in this system.
\end{abstract}


\maketitle

Macroscopic magnetic ordering phenomena in organic
materials has been one of the most exciting and interesting subjects
in physics, chemistry and materials science. Organic magnetic materials
have attracted the interest of scientists not only because of their
fundamental properties \cite{makarova-palacio}, but mainly because of their potential
applications in \textit{high-tech} devices. Among them, we have to mention
applications in medicine and biology, where organic magnetic
materials would be highly desirable due to inherent biocompatibility 
\cite{salata}.

Magnetism has been observed recently in
graphite samples \cite{fernando, esquinazi, kopelevich} and its origin
is attributed to the creation of vacancies in the material, which would
lead to the appearance of unpaired spins and consequent magnetic moments.
Previous studies \cite{rieminem-prl, rieminem-njp} have
confirmed that single-atom vacancies indeed lead to the appearance of
a center with a magnetic moment. A recent study has shown that the existence of
magnetism in vacancies induced by the removal of more than one atom can
lead to non-magnetic states, as well as magnetic ones \cite{carlsson}.

Vacancy creation in graphite samples can be achieved 
either by ion irradiation \cite{rieminem-prl} or
chemical modification \cite{fernando}. After almost two years of the
production of magnetic graphite samples by some of the present authors
using the chemical modification method \cite{fernando}, its magnetism
remains just as strong as before, at room temperature and air exposure.
It is natural, then, to ask what happens when a
foreign species is adsorbed on the magnetic center. In this Letter,
we analyze how the magnetic moment induced by single-atom vacancies in graphene
(SAV-g) changes when a single atom of H, N or O is adsorbed on the
magnetic center. As will be shown, the adsorption of these atomic
species changes the fundamental mechanisms through which the magnetic
moment is created in very different and surprising ways.

Our calculations have been performed in the framework of Density
Functional Theory \cite{kohn,KS} using the program SIESTA \cite{siesta}.
We have used Troullier-Martins pseudopotentials \cite{tm}, including
nonlinear core corrections for N, the PBE-GGA exchange-correlation
functional \cite{pbe}, and a split-valence double-zeta basis set with
polarization orbitals (DZP) for all atoms. Because the Fermi surface
of graphene is a single point, Monkhorst-Pack grids \cite{mpack}
of 9x9x1 and 5x5x1 points were used for atomic relaxations in the small
(4x4x1, SSC) and large (8x8x1, LSC) supercells, respectively, which amount to
equivalent k-point samplings of 36x36x1 and 40x40x1 in the Brilluoin Zone (BZ) of a unit cell of
graphene. Densities of states were calculated with denser grids,
equivalent to BZ samplings of 96x96x3 and a gaussian smearing of 0.08 eV. The height of the supercells
was fixed to $c/a=4.2$ in the case of SSC, and $c/a=8.2$
for LSC. All forces on atoms were smaller than 0.02 eV/\AA, and
cell relaxations included an additional constraint requiring that 
pressures be smaller than 0.05 GPa. Throughout the text, all net spin charges, 
$\delta \sigma = \sigma_{\uparrow} - \sigma_{\downarrow}$,
are referred to in units of the electron spin, that is, $\sigma_e=\frac{1}{2}$, 
and net spin densities, $\delta\rho_{\sigma}(\vec r)=\rho_{\uparrow}(\vec r)-
\rho_{\downarrow}(\vec r)$ are in units of $\delta \sigma /$Bohr$^3$.

\begin{figure}[htb]
\begin{center}
\includegraphics[width=0.45\textwidth]{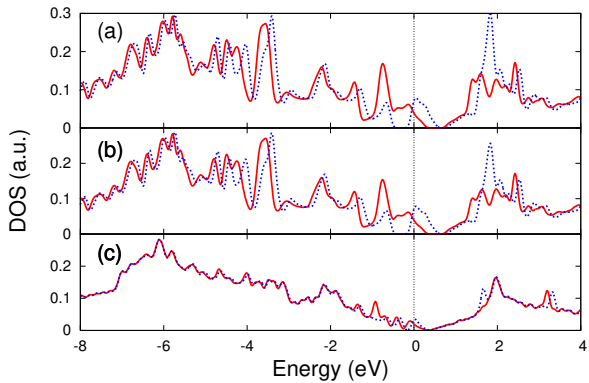}
\caption{(Color online) (a) Spin-dependent TDOS, per atom, for SSC of bare SAV-g.
(b) Spin-dependent PDOS of all C atoms of SSC with bare SAV-g. (c) TDOS of LSC
of bare SAV-g. Red, solid lines: spin up. Blue, dashed lines: spin down.}
\label{fig1}
\end{center}
\end{figure}

For both SSCs and LSCs, simple relaxation of atomic coordinates with
of bare SAV-g, simple relaxation of atomic coordinates with 
supercell vectors parallel to those of a unit cell of perfect graphene yields
stress tensors with non-zero $xy$ and $yx$ elements with magnitude comparable 
to that of the $xx$ component; furthermore, the $xx$ and $yy$ components have
very different magnitudes. 
Therefore, there is room for further relaxation of the supercell.
Variable-cell relaxation yields lattice vectors with
$|\vec a_1| \neq  |\vec a_2|$ and $\theta=61.5^\circ$ for a bare SAV-g SSC. For
the bare SAV-g LSC, we have $|\vec a_1| =  |\vec a_2|$ and $\theta=59.5^\circ$,
which shows that the in-plane interaction between vacancies plays
an important role in the process of changes of supercell symmetry. The
spin-dependent total density of states (TDOS) per C atom has two peaks close to the
Fermi level, as shown in Figs. \ref{fig1}(a) and (b), but it is clear that the peaks for the
SSC TDOS are higher and farther from each other than the ones in the LSC
TDOS. It can also be seen from the 2p projected density of states (PDOS)
per C atom (Fig. \ref{fig1}(c)) that the TDOS features are predominantly determined
by those states. Therefore, in agreement with earlier DFT
calculations \cite{carlsson}, we find 2p states to be the responsible by the
observed magnetism in bare SAV-g.

\begin{figure}[htbp]
\begin{center}
\includegraphics[width=0.35\textwidth]{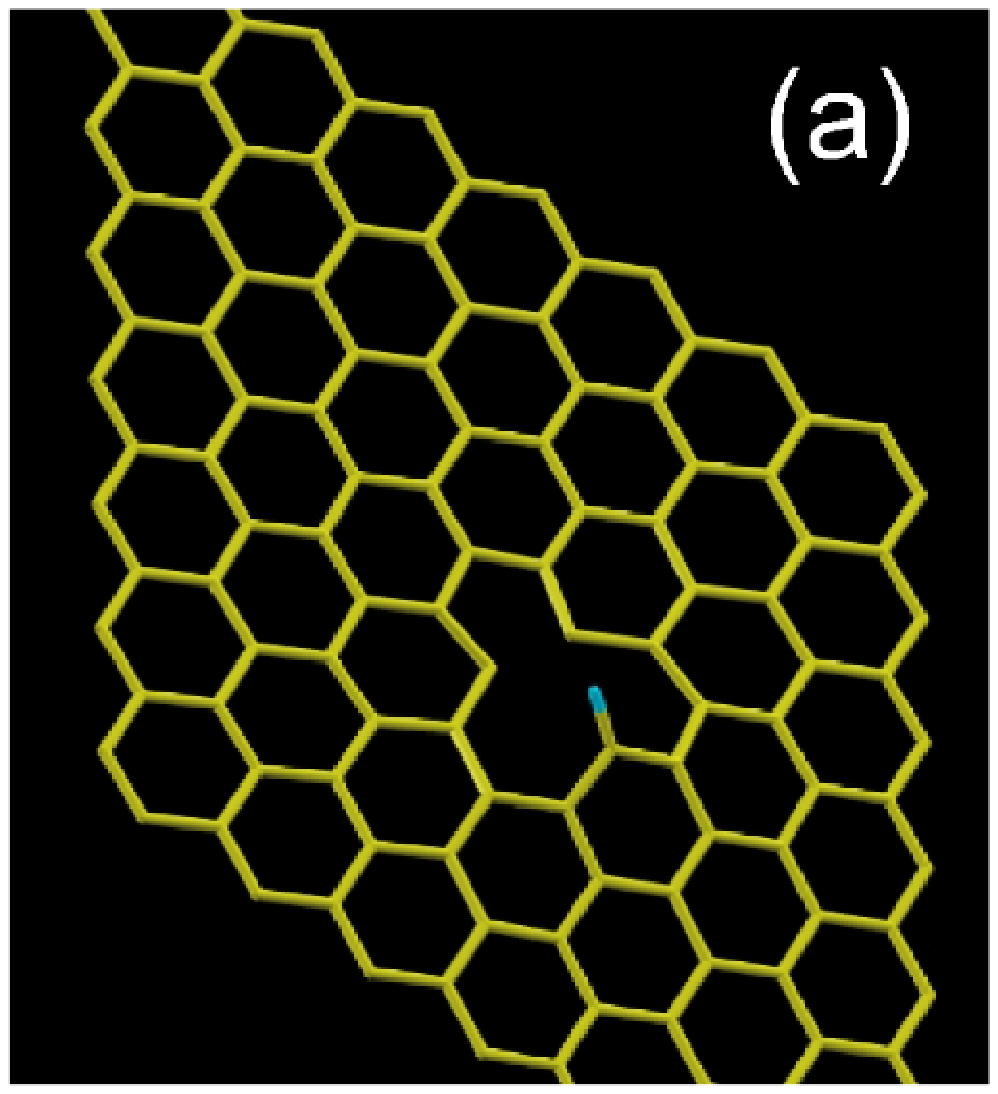}
\includegraphics[width=0.35\textwidth]{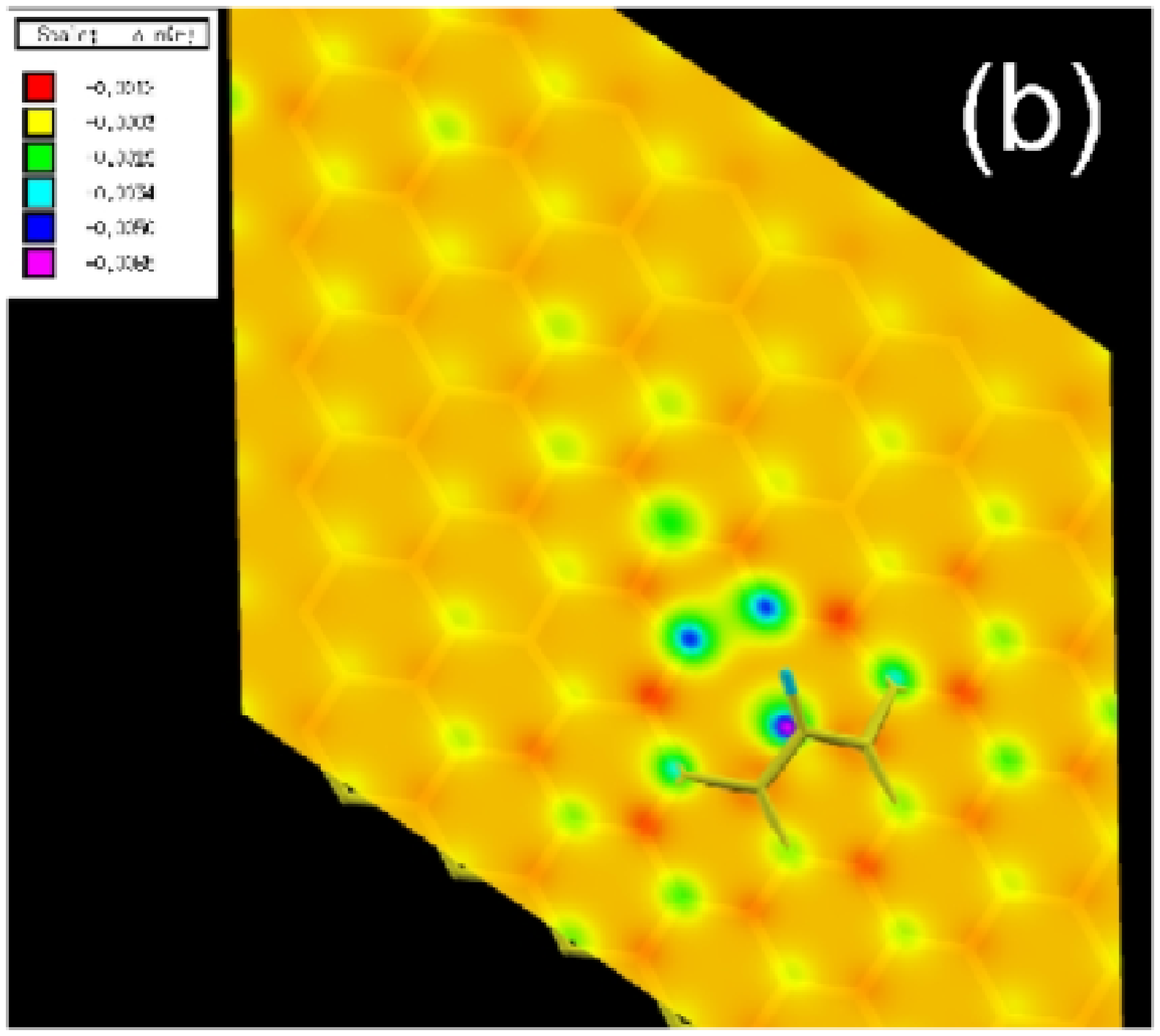}
\caption{(Color online) (a) Relaxed SSC of H-SAV-g. (b) Liquid spin density of
SSC for supercell vectors parallel to those of a perfect graphene sheet.}
\label{fig2}
\end{center}
\end{figure}

Denoting the SAV-g with a certain adsorbed atomic species X by X-SAV-g, we
start by analyzing H adsorption on the magnetic center of a SAV-g cell.
Both the SSC and the LSC are allowed for complete relaxation, undergoing a
similar symmetry breaking of the unit cell as in the case for SAV-g. The
net magnetic moment is zero, in agreement with the results of Lehtinen
 \textit{et al}\cite{rieminem-prl}. The H atom undergoes an out-of-plane
displacement, and for the SSC the overall distortion of the sheet is very pronounced,
while it is smaller in the LSC. This suggests that relief of
mechanical tensions by cell-symmetry breaking are rather important, which
and for the SSC of a H-SAV-g with for supercell vectors parallel to the ones
of the perfect graphene unit cell, the stress tensors also have nonzero $xy$ 
and $yx$ components, as well as different $xx$ and $yy$ elements. However, the
surprising fact is that \textit{under applied mechanical tension, a liquid
spin appears in both supercells}, being $\delta \sigma=$0.94 and 0.25 for the SSC and
the LSC, respectively. This fact suggests that mechanical tension is indeed the main
responsible for this different spin, because in the LSC, there is more
room for tension distribution. Fig. \ref{fig2} shows the relaxed structure 
and the net spin charge density for the H-SAV-g LSC.

\begin{figure}[htb]
\begin{center}
\includegraphics[width=0.45\textwidth]{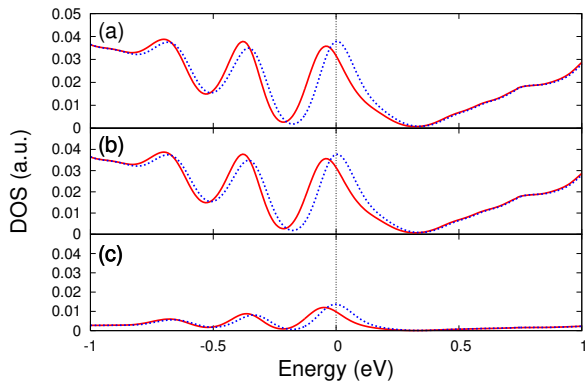}
\caption{(Color online) (a) Total spin-up and down DOS, per atom of LSC of H-SAV-g, shown
only in the vicinity of the Fermi energy (set to zero), where the two DOS
differ more significantly from each other. (b) Spin-up and spin-down 2p PDOS,
per C atom, in a LSC of H-sav-g. (c) PDOS of 1S states of the adsorbed
H atom in the LSC of H-sav-g. Red, solid lines: spin up. Blue, dashed lines: spin down.}
\label{fig3}
\end{center}
\end{figure}

While a more detailed determination of this mechanism demands a deeper
investigation, an analysis of the PDOS and spin-difference charge density
distribution, $\delta\rho_{\sigma}(\vec r)$, shows that it
involves the polarization of C atoms, while the H states
contribute with no spin polarization. The spin-dependent 2p PDOS per C atom,
as depicted in Fig. \ref{fig3}(b), clearly shows 
that the spin polarization is completely onto
the atoms of this species, while the 1s states of the H atom, shown in Fig.
\ref{fig3}(c), do not contribute to the magnetism in a significant way,
since its contribution is about four times smaller than the average
contribution of the C atoms. Moreover, in contrast to the case of bare 
SAV-g, in which the magnetic moment is concentrated mostly on the C
atom onto which the foreign atom is adsorbed, large magnetic moments 
also on the two other atoms which approach to form a pentagon. 
Even more interesting is the fact that the ordering of the spin 
polarization on the surrounding atoms has an alternating disposition,
but with decaying intensity as we move further away from the vacancy.
Since the system exhibits a net magnetic moment due to the imbalance
between spin up and spin down polarizations this arrangement is
ferrimagnetic.

\begin{figure}[htb]
\begin{center}
\includegraphics[width=0.4\textwidth]{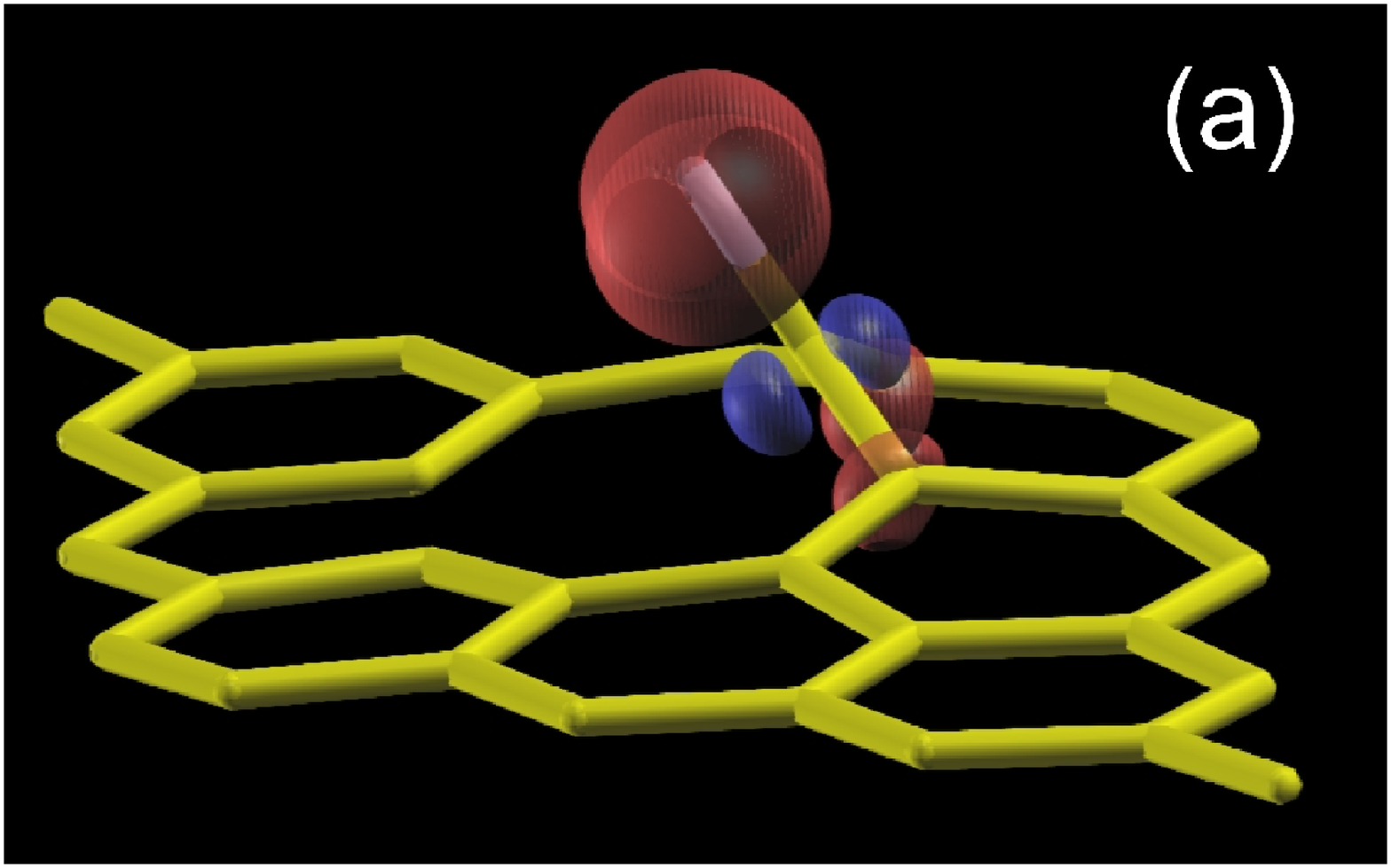}
\includegraphics[width=0.4\textwidth]{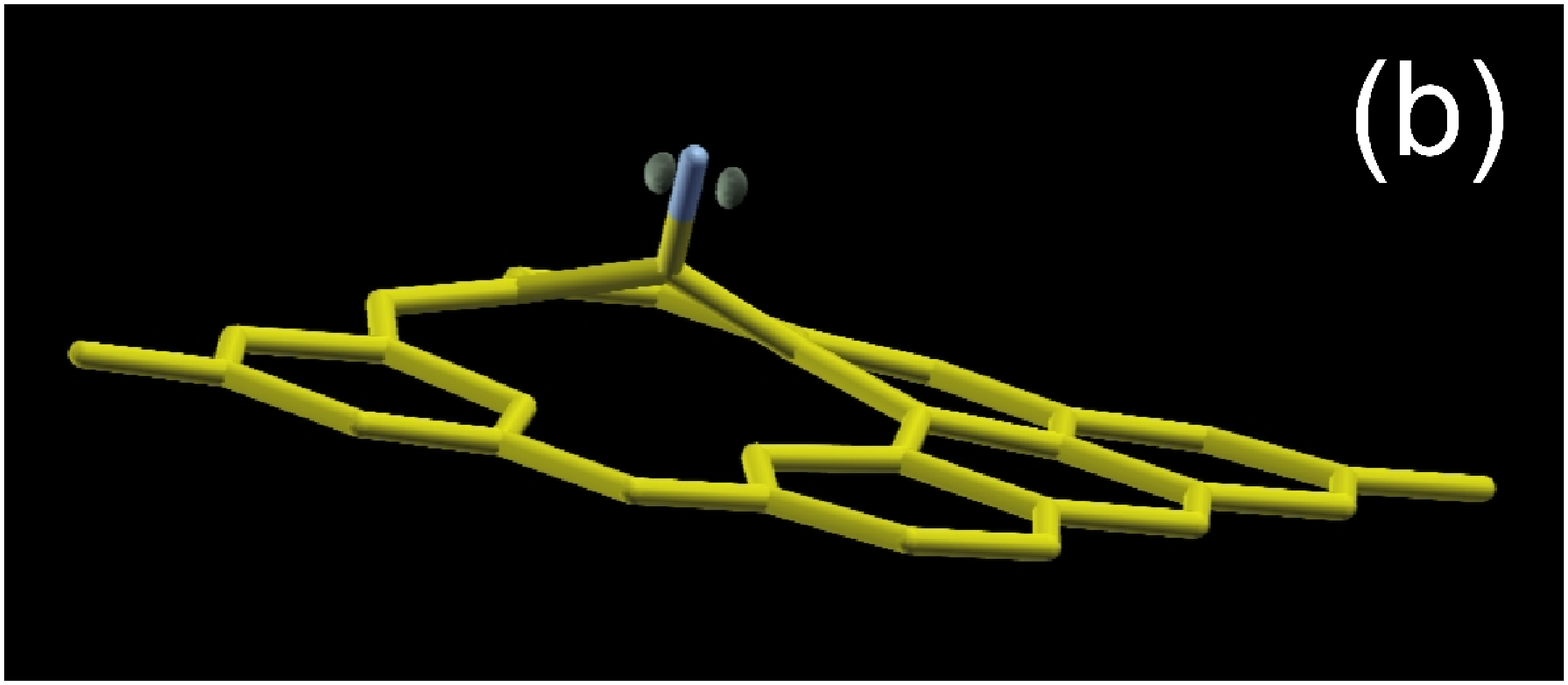}
\caption{(Color online) (a) Isosurface of $\delta\rho_{\sigma} =
\pm 0.01$ in SSC with N-SAV-g; positive $\delta\rho_{\sigma}$
are depicted in red, negative ones in blue. The widest surface is
around the N atom, and C atoms display a far smaller isosurface. (b) Isosurface
of $\delta\rho_{\sigma} = 0.2$. No negative values are observed on
atoms other than C.} \label{fig4}
\end{center}
\end{figure}

Next, we analyze the influence of N adsorption onto the magnetic center of a bare SAV-g,
in the geometry depicted in Fig. \ref{fig4}(a). Both
SSC and LSC, whether cell-relaxed or not, display $\delta \rho_{\sigma}=1$, that is, the
magnetic moment in N-SAV-g is independent of geometrical details. To better understand this,
we plot spin-difference charge density isosurfaces
with different values to check how liquid spin is distrbuted in the supercell.
For a spin-difference of $\delta \rho_{\sigma}=\pm 0.01$, we can see that the positive
spin charge is widely distributed around the adsorbed N atom, while the negative
spin charge is narrowly distributed around the closest C atom. Isosurfaces
of $\delta \rho_{\sigma}=0.2$ are tightly bound to the N atom, while on C atoms
$\delta \rho_{\sigma}=0$. Looking at the C-N bond length,
$a_{C-N}=1.25$ \AA, we can infer that the N atom makes a double bond with the C atom. This
causes the saturation of the unpaired C spin, with $\delta \sigma=1$
coming from the N atom. Since the net spin is tightly bound to the N atom,
this explains the insensitivity of the magnetism to geometrical details, in
contrast with the H-SAV-g. Although according to our calculations the ground state
for a N atom would be that in which it enters the sheet substitutionally, the situation
depicted in Fig. \ref{fig4} could be found, experimentally, in a zigzag state 
for a vacancy where many atoms have been removed. In these large vacancies,
the N atom will not enter substitutionally and such a configuration is more 
likely to occur.

\begin{figure}[htb]
\begin{center}
\includegraphics[width=0.49\textwidth]{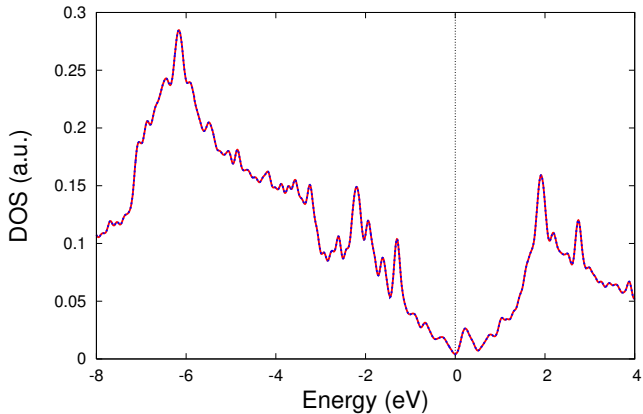}
\caption{(Color online) Total density of states per atom of O-SAV-g LSC.}
\label{fig5}
\end{center}
\end{figure}

The case of O-adsorption is somewhat the opposite of N-adsorption.
O-adsorption on the magnetic center causes the magnetic moment to completely vanish,
no matter if the cell is distorted or not, for both SSC and LSC. Although this is
in striking constrast with the results of Esquinazi \textit{et al} \cite{kopelevich},
where the authors find that the presence of O$_2$ triggers a magnetic moment, it is
important to bear in mind that the present calculations do not correspond to the
experimental situation, in which the authors use molecular oxygen. Further theoretical
work, with systems which approximate more realistically the experimental situation -
that is, layered graphite with adsorbed and perhaps interstitial O$_2$ as well - is
necessary to clarify the situation. Fig. \ref{fig5} shows the TDOS, per atom of the LSC, 
of O-SAV-g.

Our findings certainly will have important implications in most of the
applications of carbon-based magnetic materials in
nanotechnology, such as sensors, detectors, actuators, telecommunications,
electronics, biosensors, magnetic materials separation, etc.. However, those
implications will be particularly important for applications in
medicine, specially for targeted drug delivery by use of magnetic
field gradients \cite{salata, kouassi}. Our calculations imply that, if magnetic graphene is to be used in
this kind of application, one must avoid the saturation of the magnetic centers,
as much as possible, by oxygen, and that a good situation might be that in
which the magnetic centers have N adsorbed. Experimentally, it is known that 
biologically relevant molecules such as cholesterol oxidase attach to magnetic
Fe$_3$O$_4$ nanoparticles through amino and carbodiimide-activated carboxyl
groups \cite{kouassi}, both of which contain N. This suggests that proteins
could attach to magnetic graphene without the use of a ligand, which would
make the use of magnetic graphene extremely attractive for biomedical 
applications. Our results also imply that, if graphene with vacancies 
should be used for any kind of application involving its magnetic properties, 
it should be, as much as possible, oxygen-free. If H should be present in 
the system, then deposition of the graphene sheet onto a substrate might 
provide the tension necessary for the appearance of a magnetic moment due 
to lattice mismatch. The use of multi-layered graphitic structures is, in
principle, also not discarded: preliminary FPLAPW results show that bulk 
graphite with vacancies on different layers, and in different concentrations,
also exhibit a net magnetic moment \cite{to-be-published}.

In summary, we have studied three very different mechanisms that lead to 
the appearance of a net magnetic moment on a
graphene sheet with a single-atom vacancy, both bare and with foreign
atomic species adsorbed on top of it, particularly H, N and O. 
Both small and large supercells have been studied. A strong breaking of
the supercell symmetry is observed for all systems. While the magnetic moment
initially present in the bare SAV-g, regardless of the geometrical details 
of the supercell, is completely quenched by O adsorption and is remarkably
stable upon N adsorption, we find that it can be induced by tensile stress 
of the lattice when H is adsorbed in the system. Although we have studied only the 
adsorption of atomic species on the system, it is highly possible that adsorption 
of molecules is also capable of preserving the magnetism in the system,
and its implications for device manufacturing make this a good subject for further studies.

\begin{center}
  {\bf ACKNOWLEDGMENTS}
\end{center}

R. Y. Oeiras and F. M. Ara\'ujo-Moreira acknowledge brazilian funding
agencies CAPES, FAPESP  and CNPq for financial support. A. W. Mombr\'u
acknowledges uruguayan funding agencies, PEDECIBA, CSIC and CONICYT. 
M. Ver\'{\i}ssimo-Alves
acknowledges The Abdus Salam ICTP for the grant of a post-doctoral
fellowship, Drs. J. M. Carlsson and
M. Scheffler for a preprint on related work, and Dr. R. B. Capaz, Dr. A. J. R.
da Silva, Dr. S. Reich, Dr. A.
A. Leit\~ao and Dr. E. Tosatti for enlightening discussions.
Most of the calculations were performed using the CINECA,
Bologna, Italy, and the CENAPAD-SP, Campinas, Brazil, computational facilities.

\newpage

\end{document}